\documentclass[a4paper,11pt]{article}
\usepackage{pos}

\title{Status update of the MACE Gamma-ray telescope}
 \ShortTitle{MACE telescope}

\author*[a,b]{K. K. Yadav}

\affiliation[a]{Astrophysical Sciences Division, Bhabha Atomic Research Centre,\\
  Mumbai 400094, India}

\affiliation[b]{Homi Bhabha National Institute,\\
  Mumbai 400094, India}

\forColl{HiGRO} 

\emailAdd{kkyadav@barc.gov.in}

\abstract{MACE (Major Atmospheric Cherenkov Experiment), an imaging atmospheric Cherenkov telescope, has recently been 
installed by the HiGRO (Himalayan Gamma-Ray Observatory) collaboration at Hanle (32.8$^\circ$N, 
78.9$^\circ$E, 4270m asl) in Ladakh region of North India. The telescope has a 21m diameter large light collector 
consisting of indigenously developed 1424 square-shaped diamond turned spherical aluminum mirror facets of size 
$\sim$ 0.5m$\times$0.5m. MACE is the second largest Cherenkov telescope at the highest altitude in the northern hemisphere. 
The imaging camera of the telescope consists of 1088 photo-multiplier tubes with a uniform pixel resolution 
of $\sim 0.125^\circ$ covering a field of view of $\sim$ 4.0$^\circ$ $\times$ 4.0$^\circ$. The main objective of the 
MACE telescope is to study gamma-ray sources mainly in the unexplored energy region 20 -100 GeV and beyond with high 
sensitivity. In this paper, we describe the key design features and current status of MACE including results from 
the trial observations of the telescope.}

\FullConference{37$^{\rm{th}}$ International Cosmic Ray Conference (ICRC 2021)\\
		July 12th -- 23rd, 2021\\
		Online -- Berlin, Germany}

\begin{document}
\maketitle
\section{Introduction}
MACE (Major Atmospheric Cherenkov Experiment), a very high energy (VHE) $\gamma$-ray telescope based on imaging atmospheric 
Cherenkov technique, is currently under commissioning at Hanle, India \cite{Koul2017,Singh2021}. 
The Indian astronomical site at Hanle (32.8$^\circ$N, 78.9$^\circ$E, 4270 m asl) in the Himalayan range is 
at the highest altitude in the world for any existing imaging atmospheric Cherenkov telescope (IACT) at present.  
The main objective of the MACE telescope is the observation of the VHE $\gamma$-ray photons with energy 30 GeV and above, 
which is traditionally considered as the window for space-based telescopes. Observations with the MACE telescope will 
help in understanding the production and propagation of energetic $\gamma$-ray photons in the non-thermal Universe and 
to also explore some phenomena which are not accessible to terrestrial accelerators. Some of the important astrophysical 
motivations for the MACE telescope include: discovery of new $\gamma$-ray sources, identification of unassociated sources 
in the high energy \emph{Fermi} catalog, detection of the $\gamma$-ray emission from various types of known sources 
(active galactic nuclei, gamma-ray bursts, pulsars and so on) at different ages of the Universe, providing evidence for 
the enigmatic jet formation process etc. Beyond probing the non-thermal Universe and cosmic accelerators, the VHE observations 
using the MACE telescope are also expected to address a range of cosmological topics such as measurement of the 
intensity of extragalactic background light, cosmic ray electron spectrum, search for nature of dark 
matter candidates like weakly interacting massive particles \cite{Aharonian2008,Hillas2013,Giuseppe2019}. 

\section{The MACE Telescope: Salient Features}
\begin{center}
\begin{figure}
\includegraphics[width=0.5\textwidth]{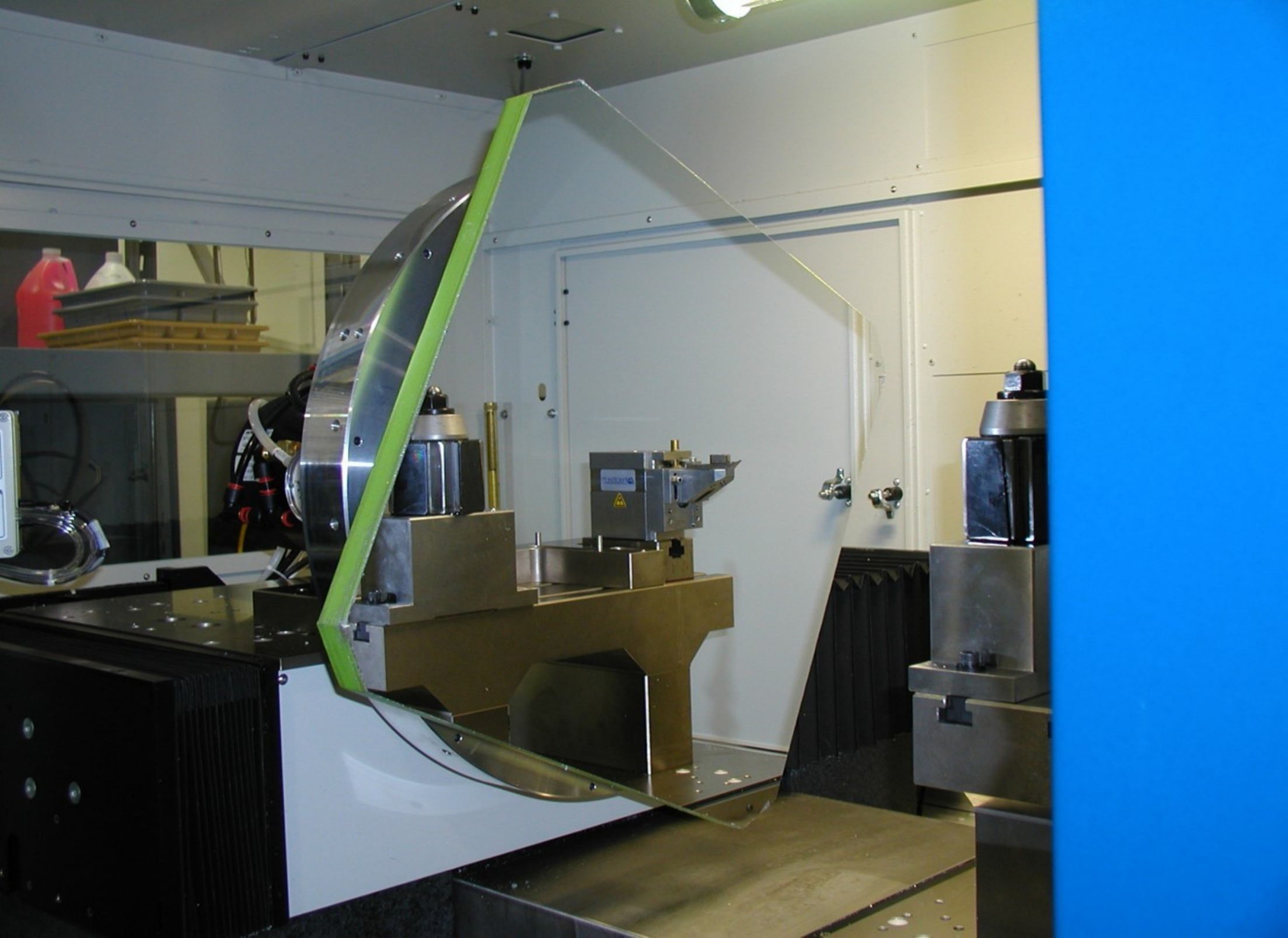}
\includegraphics[width=0.5\textwidth]{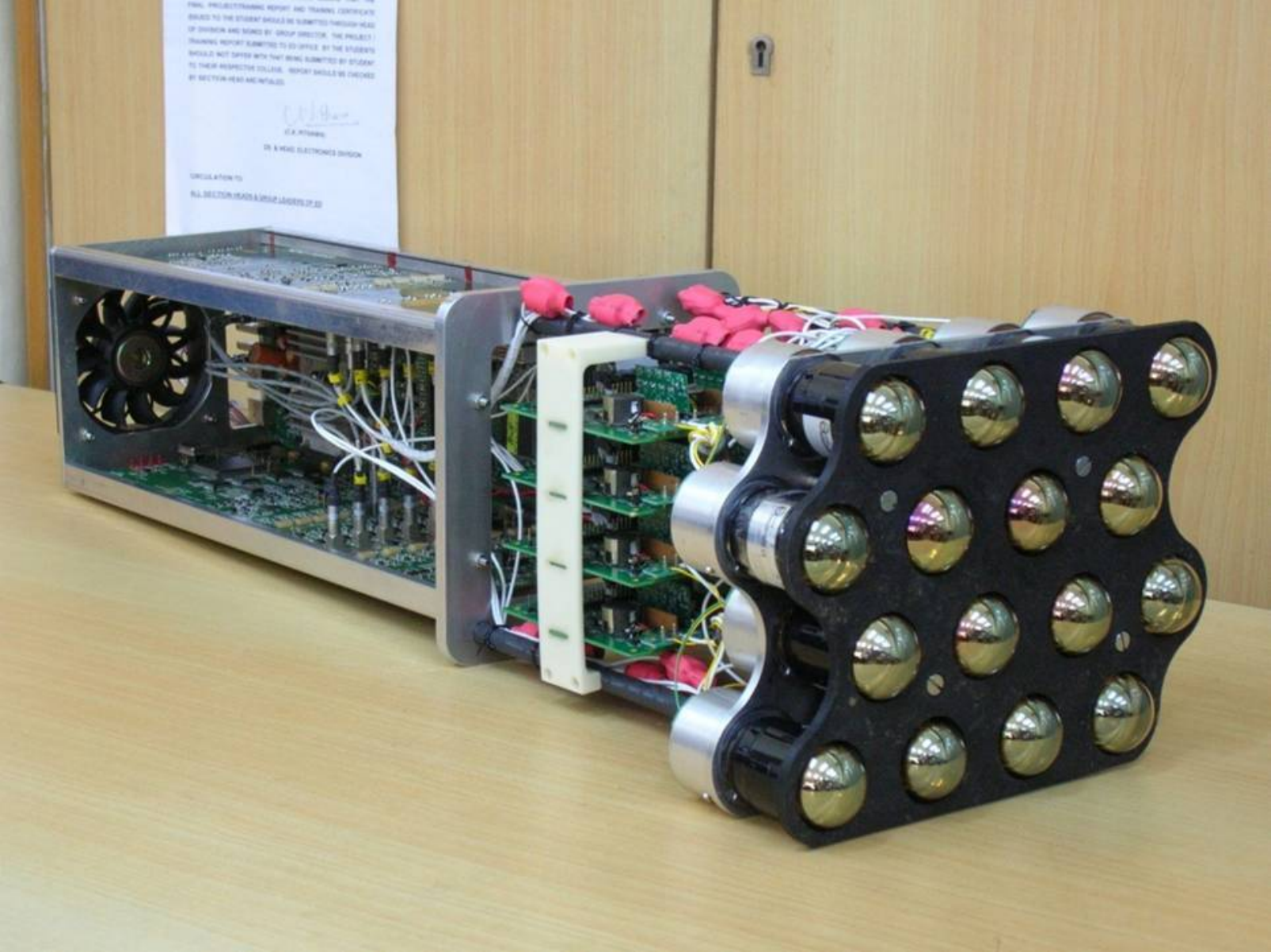}
\caption{Left: A diamond turned mirror facet for the MACE light collector. 
        Right: 16-channel camera integrated module for the camera of MACE telescope.}
\label{mirror+cim}
\end{figure}	
\end{center}

The MACE telescope is equipped with a 21m diameter quasi-parabolic reflector with $f/d$ ratio of $\sim$ 1.2. 
The quasi-parabolic design of such a large reflector helps in reducing the optical aberrations of the telescope. 
To achieve such a large light collector area, reflector of the MACE telescope is segmented into 1424 small, 
square-shaped spherical mirror facets of size 0.488m $\times$ 0.488m each with varying focal lengths (Figure \ref{mirror+cim}(Left)). 
Four such mirrors with similar focal length are mounted on a single panel of size 0.986m $\times$ 0.986m each 
and a total of 356 mirror panels with varying focal length in the range 25m to 26.2m are obtained. 
All the 356 mirror panels mounted on the telescope basket are manually aligned in such a way that the 
resulting surface behaves like a single quasi-parabolic reflector with focal length gradually increasing 
from the center of the basket towards the periphery. An active mirror alignment control system is used for 
orienting each mirror panel to achieve the desired optical quality of the reflector. The metallic mirror facets 
are made up of Aluminium alloy supported by a honeycomb structure. Diamond turning technology \cite{Sugano1987} 
has been used to fabricate all the mirror facets in order to achieve desired mirror quality in terms of surface 
finish and accuracy within the country (Figure \ref{mirror+cim}(Left)). The reflectance of these mirror facets is greater than 85$\%$ in the 
wavelength range 280-700 nm. Due to its large size, the MACE telescope is not protected by a dome and 
the mirrors are continuously exposed to the environment. The reflecting surface of each mirror facet is coated 
with a thin (100-150 nm) layer of SiO$_2$ for protecting the reflecting surface and ensuring its longevity. 
\par
The telescope deploys a 1088-pixel imaging camera with resolution of 0.125$^\circ$ at its focal plane, covering 
a field of view of $4.36^\circ \times 4.03^\circ$. Six stage photomultiplier tubes (pixel) with 38 mm diameter arranged 
at a triangular pitch of 55 mm are used in the MACE-camera for detection of the Cherenkov photons. Each photomultiplier 
tube is provided with a hexagoanl front-coated light concentrator in order to enhance the light collection efficiency 
by collecting the Cherenkov photons incident in the dead space between adjacent pixels. The light collection efficiency 
of these compound parabolic concentrators is more than 85$\%$ in the wavelength range of 230-600 nm. 
The MACE-camera is modular in design and consists of 68 camera integrated modules (CIMs) of 16 channels each. 
Each module channel has its signal processing electronics (photomultiplier tubes with programmable high voltage generators, 
pre-amplifiers, amplifiers, discriminators, first level trigger generation logic, and signal digitization circuitry) housed 
in it (Figure \ref{mirror+cim}(Right)). An analogue switched capacitor array DRS-4 is used as ring sampler at 1 GHz speed for continuous digitization of 
the signal from the photomultiplier tubes. The signal from the photomultiplier tubes is simultaneously amplified at low and 
high gain of 14 and 140 respectively to ensure a large dynamic range. The discriminator output amplitude of each channel is 
used for monitoring its single channel rate and also for generating the first-level trigger from an individual CIM. 
The first-level triggers from all the modules are collated in a second-level trigger generator where proximity of 
the triggered pixels in adjacent modules is checked. The innermost 576 pixels (24$\times$24) are used for generating the 
trigger according to predefined logic for nearest pairs, triplets, quadruplets, etc of the pixels. After the generation 
of the second-level trigger, the data from all the 68 modules are collected by the data concentrator, which in turn sends 
them to the data acquisition computer in the control room through optical fibres. It is expected that about 50GB of data will 
be stored during every hour of observation with the MACE telescope \cite{Srivastava2020,Sarkar2020}. 
\par
An altitude-azimuth mount drive system of the MACE telescope provides the stability of 21m diameter large mechanical structure with 
180 ton weight by using a track and wheel design. Two azimuth drive wheels are coupled to three phase, permanent 
magnet brushless AC servo motors through multi-stage gearboxes for providing the azimuth motion. The elevation movement is 
provided through a gearbox coupled to a  13-section bull-gear assembly of 11.6m radius. All the drives are provided with
counter torque capability to avoid gear backlash error. The motors are driven by pulse width modulated drive amplifiers 
powered by 480 Volt DC from a solar power station. The positions of the two axes are monitored by a 25-bit absolute optical 
encoders with $\sim$ 20 arcsec accuracy. Both the azimuth and elevation gear boxes also have high-speed options to move 
the telescope at speed of 3$^\circ s^{-1}$ to quickly point the telescope in the direction of transient events like gamma-ray bursts 
in the sky. The MACE drive system can provide tracking accuracy of better than 1 arcmin in wind speeds of up to 30 km h$^{-1}$. 
The telescope is automatically brought to the parking position if the sustained wind speed is more than 40 km h$^{-1}$. 
Moreover, the telescope structure is designed for survival at a wind speed of 150 km h$^{-1}$.
\begin{center}
\begin{figure}
\includegraphics[width=1.0\textwidth]{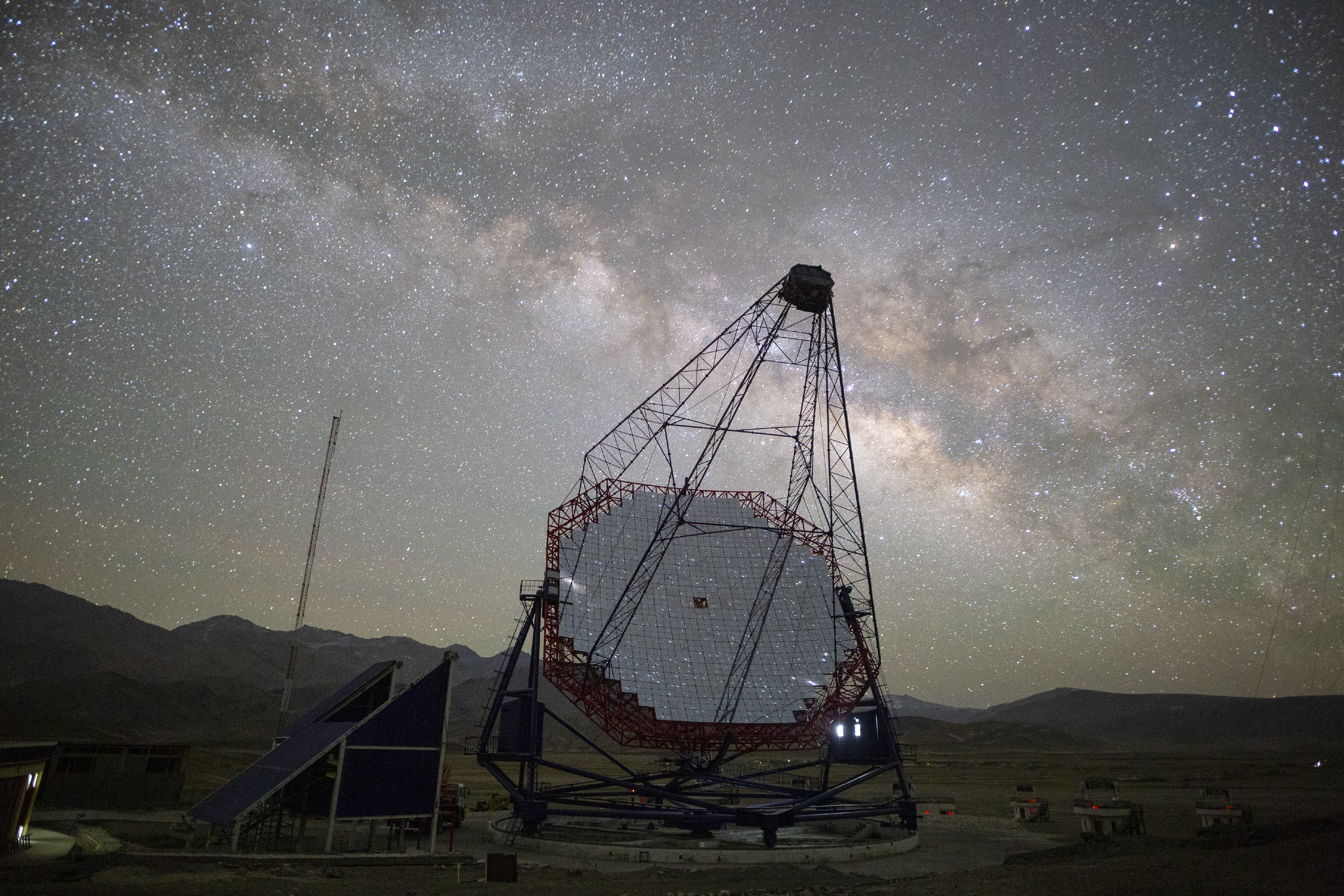}
\caption{The MACE telescope installed at Hanle-Ladakh, India.}
\label{mace}
\end{figure}	
\end{center}

\begin{center}
\begin{figure}
\includegraphics[width=1.0\textwidth]{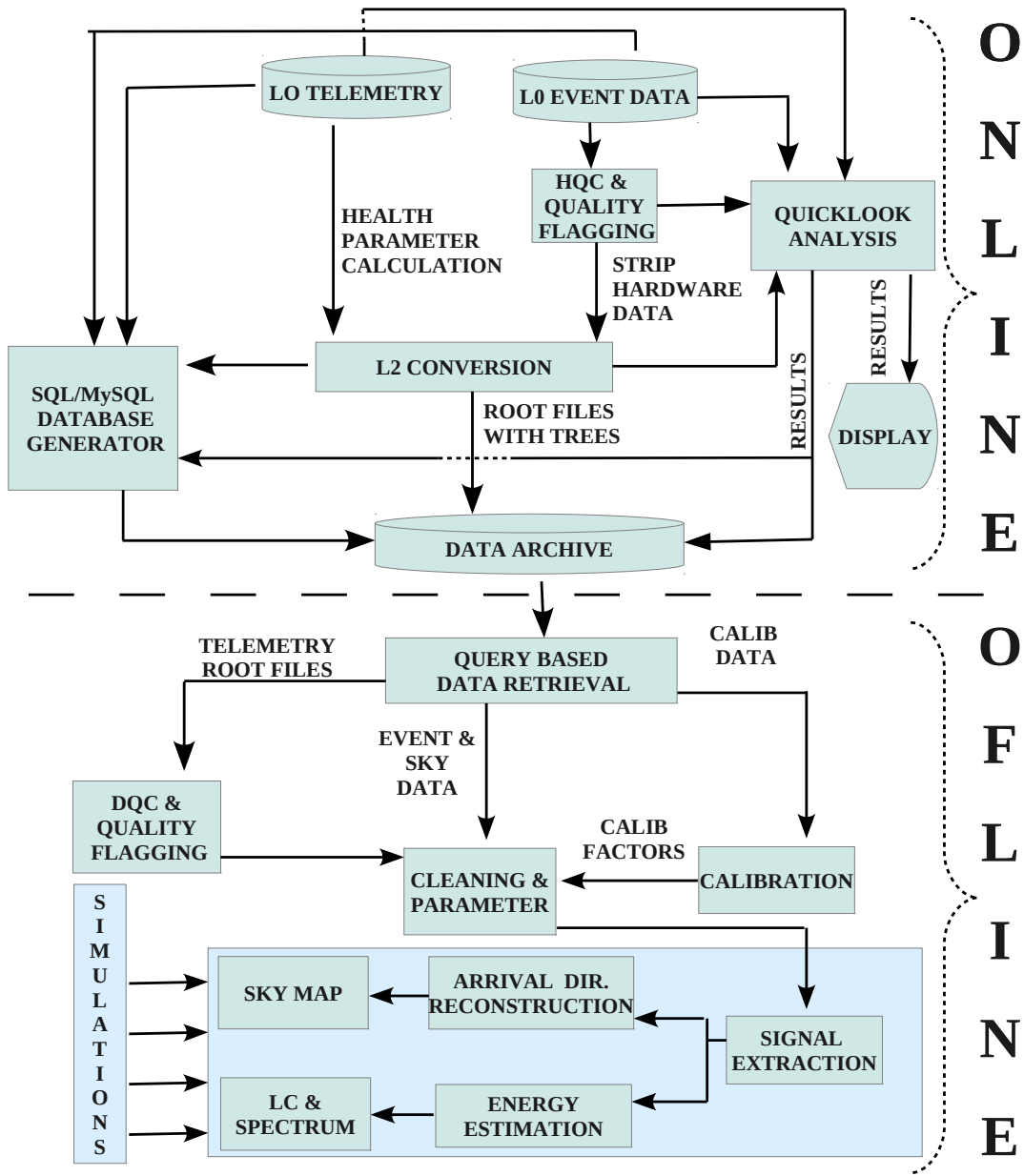}
\caption{Flowchart for the MACE data Analysis Package (MAP).}
\label{map}
\end{figure}	
\end{center}
\section{Present Status of the MACE Telescope}
The installation of the MACE telescope at Hanle site was successfully completed in October 2020. A recent picture of 
the telescope at the site is depicted in Figure \ref{mace}. The trial runs for testing the performance of different 
components of the fully assembled telescope started in November 2020. Monte Carlo simulation studies carried out 
using the CORSIKA package suggest that the MACE telescope is expected to have a $\gamma$-ray trigger energy threshold of $\sim$ 
20 GeV in the low zenith angle range below 40$^\circ$ \cite{Borwankar2016}. At high zenith angles above 40$^\circ$, 
the threshold energy increases to $\sim$ 150 GeV. The dynamic energy range of the MACE telescope is expected to be 
20 GeV-5 TeV with a total trigger rate of $\sim$ 1 kHz. The integral sensitivity for a point source with Crab Nebula like 
spectrum above 30 GeV is 2.7$\%$ at 5$\sigma$ statistical significance level in 50 hours of 
observation \cite{Sharma2017}. The angular resolution of MACE telescope is estimated as $\sim$ 0.21$^\circ$ in the energy range 
of 30-50 GeV and this improves to $\sim$ 0.06$^\circ$ in the energy band 1.8-3 TeV \cite{Borwankar2020}. The telescope is expected to 
have energy resolutions of $\sim$ 40$\%$ and $\sim$ 19$\%$ in the energy bands 30-50 GeV and 1.8-3 TeV respectively.  
For an effective analysis of the real time data collected with the MACE telescope, a data analysis software called MAP (MACE Analysis Package) 
has been developed. The MAP represents collection of ROOT-based programs/routines written in C-language under LINUX platform. 
Different analysis steps included in the MAP are outlined in Figure \ref{map}. Another version of the MAP based on Python libraries is also 
under development.

\begin{center}
\begin{figure}
\includegraphics[width=0.5\textwidth]{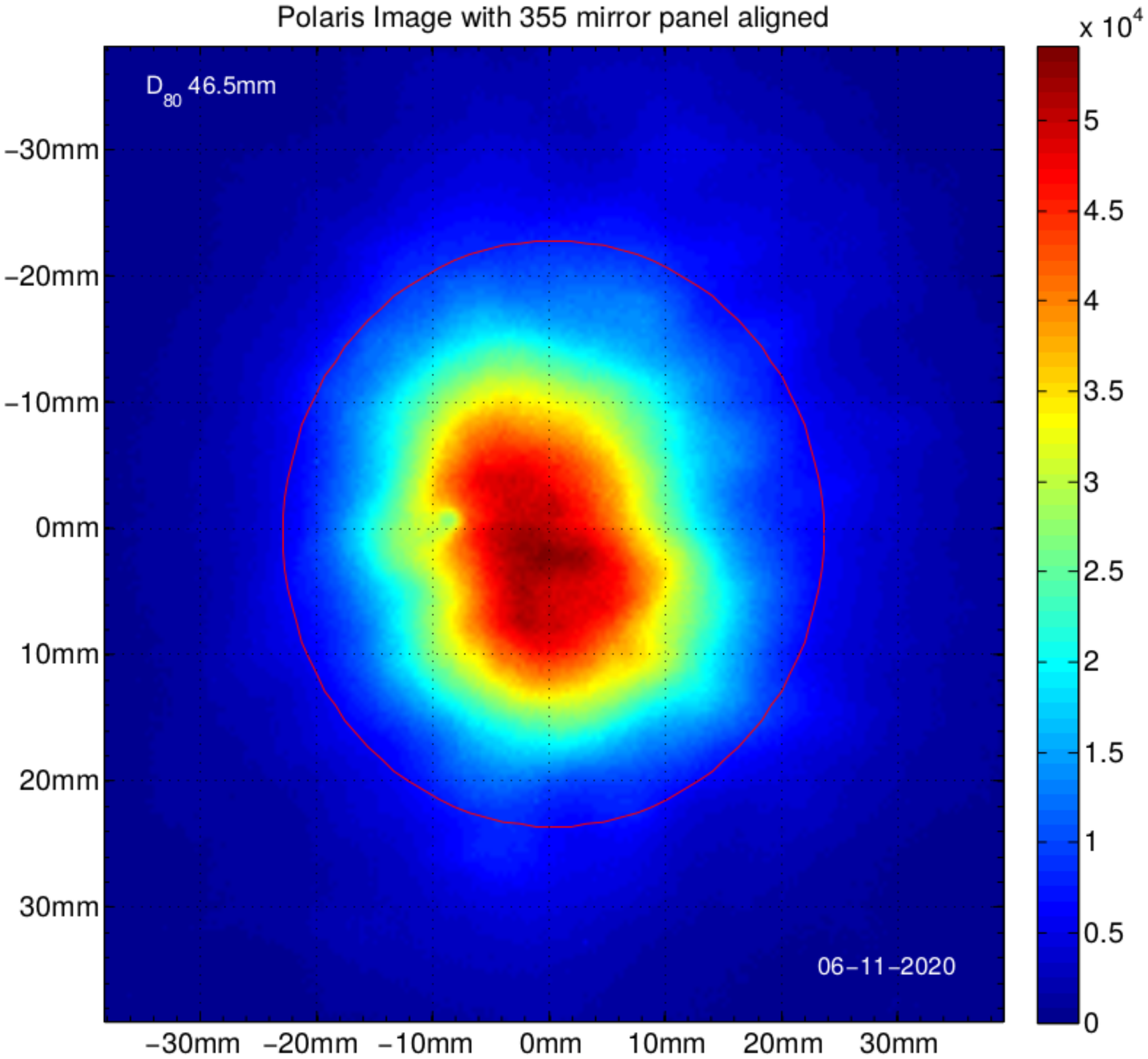}
\includegraphics[width=0.55\textwidth]{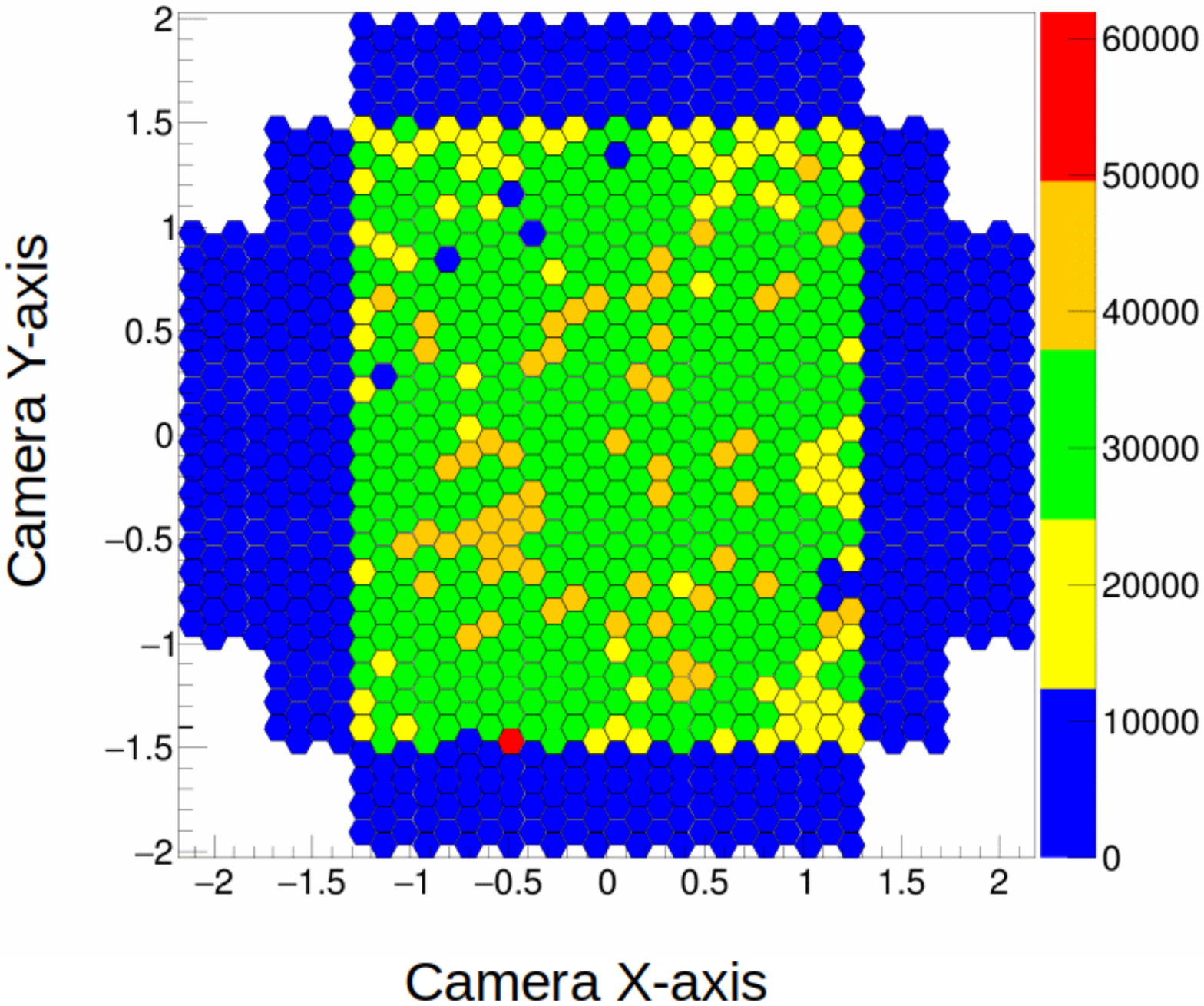}
\caption{Left: The point spread function of the MACE telescope measured for Polaris.
         Right: Hit pattern uniformity in the trigger region of the camera during the Crab Nebula observations 
        for $\sim$ 1 hour on April 1, 2021.}
\label{psf+hit}
\end{figure}	
\end{center} 

\section{Preliminary Results}
The MACE telescope, currently under its commissioning phase, has been deployed for special trail runs since October 2020. 
Alignment of the 356 mirror panels on telescope basket is carried out with the help of an active mirror alignment control 
system while tracking the pole star. A CCD camera along with 180mm focal length lens mounted on the center of the telescope 
basket is used to capture the reflected image of an optical source on telescope focal plane. The image of pole star captured by 
the CCD camera on the focal plane after background subtraction and image cleaning is shown in Figure \ref{psf+hit}(Left). 
D80 (defined as the diameter of circle, concentric with the centroid of the image, which contains 80$\%$ intensity of the image) 
of the image is found to be 46.5 mm after performing a Gaussian fit to  the pole star image data. This characterizes the point 
sperad function of the telescope. 
\par
In order to record the Cherenov event data, the information of hit channels in the camera of the telescope is also stored. The number 
of hits for each triggered pixel in the trigger region (inner 576 pixels) of the camera during the trial run in the direction of 
the Crab Nebula on the night of April 1, 2021 for about 1 hour of observation is reported in Figure \ref{psf+hit}(Right). This 
indicates the homogeneity of the camera of the MACE telescope in generating the trigger in response to the Cherenkov events.
\par
\begin{center}
\begin{figure}
\includegraphics[width=1.0\textwidth]{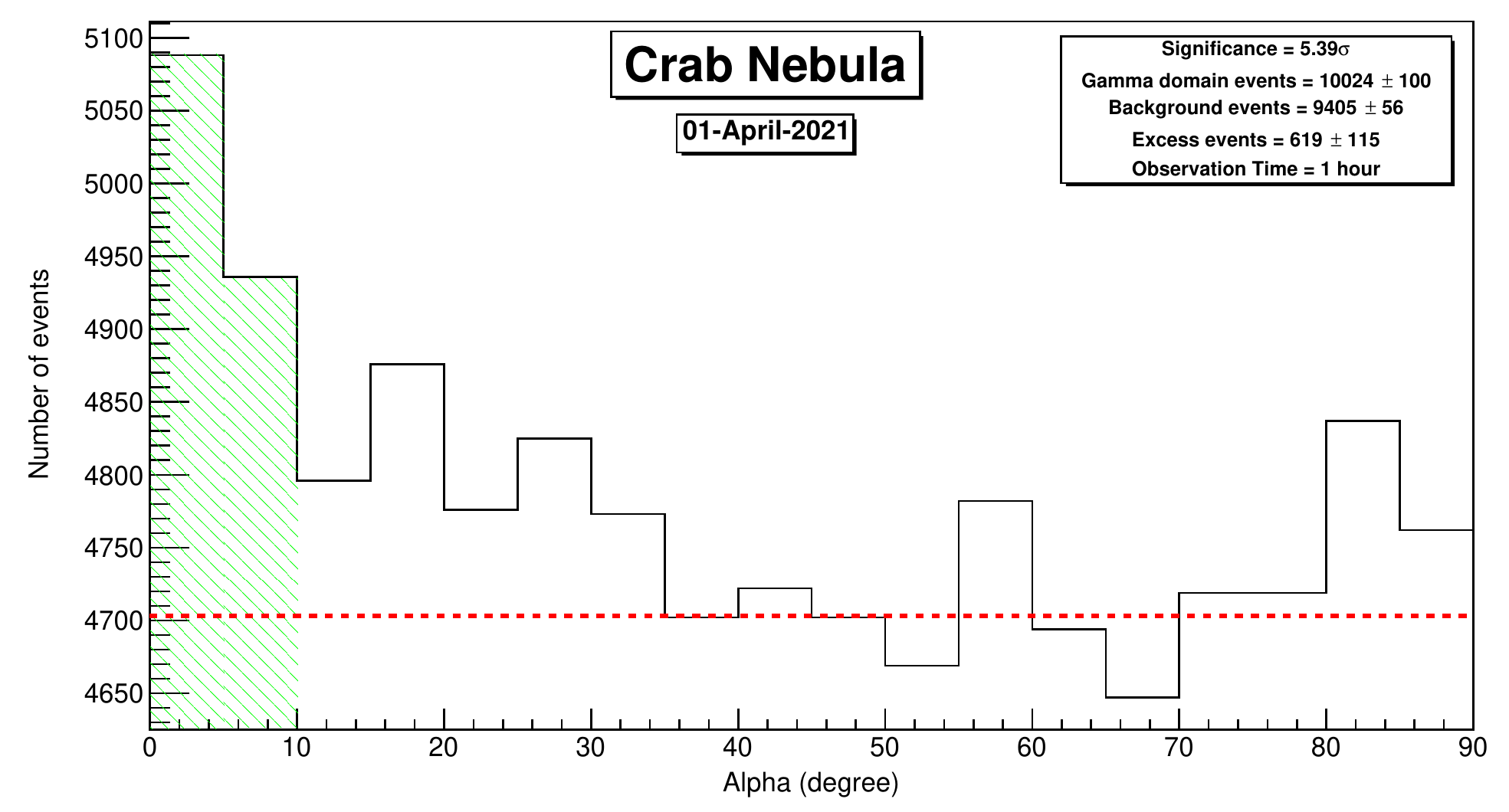}
\caption{Preliminary results from the Crab Nebula observations on April 1, 2021 for a live time of $\sim$ 1 hour.}
\label{alpha}
\end{figure}	
\end{center}

Preliminary analysis of the data collected on the Crab Nebula on April 1, 2021 has been performed using the standard 
Hillas Parameterization method \cite{Hillas1985} implemented in the MAP. The distribution of the Hillas orientation parameter ($\alpha$) for 
the recorded Cherenkov events is depicted in Figure \ref{alpha}. The $\alpha$-plot shown here only represents the initial 
performance of the MACE telescope without standard data quality checks for IACT observations. Although there is an indication of 
the presence of $\gamma$-ray like events in the source region at $\sim$ 5.4$\sigma$ statistical significance level in Figure \ref{alpha}, 
yet these preliminary results in the present form cannot be referred to as the first light of the MACE telescope. A detailed analysis of 
the data collected on the Crab Nebula with MACE in the month of April 2021 is currently underway and the final results of 
the first light are expected very soon.

\section{Summary}
A group of very high energy $\gamma$-ray astronomers are actively involved in setting up the MACE telescope 
in India under HiGRO (Himalayan Gamma Ray Observatory) Collaboration. The construction and installation 
of the telescope at Hanle-site has been successfully completed. Currently, the MACE telescope is under its 
commissioning phase and its first light is expected very soon. Subsequently, the telescope will enter 
into the science operatione phase. Once fully operational, the MACE telescope, equipped with a 21m diameter light 
collector and 1088-pixel camera, will be the second largest $\gamma$-ray telescope (after the Large Size Telescope prototype (LST-1) 
at the Cherenkov Telescope Array) in the northern hemisphere. However, altitude of the MACE-site is the highest in the world. 
This geographical advantage along with the large light collector will help in exploring the $\gamma$-ray sky above 
15GeV and also provide an excellent energy overlap with the space-based $\gamma$-ray observatories.
\section*{Acknowledgements}
The HiGRO collaboration thanks all the colleagues from various Divisions at Bhabha Atomic Research Centre, who have 
been involved at different stages of the construction and installation of the MACE telescope at Hanle, India. 


\clearpage
\section*{Full Authors List: \Coll\ Collaboration}
%
%
\noindent
N. Bhatt$^1$, S. Bhattacharyya$^{1,2}$, C. Borwankar$^1$, K. Chanchalani$^1$, P. Chandra$^1$, V. R. Chitnis$^3$, 
N. Chouhan$^1$, M. P. Das$^1$, V. K. Dhar$^1$, B. Ghosal$^1$, S. Godambe$^1$, S. Godiyal$^1$, K. K. Gour$^1$, J. Hariharan$^1$, 
M. Khurana$^1$, M. Kothari$^1$, S. Kotwal$^1$, M. K. Koul$^1$, N. Kumar$^1$, N. Kumar$^1$, C. P. Kushwaha$^1$, N. Mankuzhiyil$^1$, 
P. Marandi$^1$, K. Nand$^1$, S. Norlha$^1$, D. Sarkar$^1$, M. Sharma$^1$, K. K. Singh$^1$, R. Thubstan$^1$, 
A. Tolamatti$^1$, K. Venugopal$^1$, K. K. Yadav$^{1,2}$\\

%
\noindent
$^1$Astrophysical Sciences Division, Bhabha Atomic Research Centre, Mumbai 400094, India.\\
$^2$Homi Bhabha National Institute, Anushakti Nagar,Mumbai 400094, India.\\
$^3$Department of High Energy Physics, Tata Institute of Fundamental Research, Homi Bhabha Road, Colaba, Mumbai 400005, India.

\end{document}